\documentclass[runningheads,a4paper]{llncs}

\usepackage[T1]{fontenc}
\usepackage{graphicx}
\usepackage{multirow}
\usepackage{xcolor}
\newcommand{\revC}[1]{{#1}}
\newcommand{\revD}[1]{{#1}}
\usepackage{array}
\usepackage{booktabs}
\usepackage[final]{microtype}
\usepackage{amsmath}
\usepackage{amssymb}
\usepackage{tikz}
\usetikzlibrary{shapes.geometric,shapes.misc,arrows.meta,positioning,fit,backgrounds,calc}
\usepackage[hidelinks]{hyperref}

\AtBeginDocument{%
  \setlength{\textfloatsep}{6pt plus 2pt minus 2pt}%
  \setlength{\intextsep}{6pt plus 2pt minus 2pt}%
  \setlength{\floatsep}{6pt plus 2pt minus 2pt}%
  \setlength{\abovecaptionskip}{4pt}%
  \setlength{\belowcaptionskip}{2pt}%
  \raggedbottom%
}

\begin{document}

\title{CAPRA: Scaling Feedback on Software Architecture Deliverables with a Multi-Agent LLM System}
\titlerunning{CAPRA: Scaling Feedback on Software Architecture Deliverables}

\author{Marco Becattini\inst{1} \and
Niccol\`{o} Caselli\inst{1} \and
Matteo Minin\inst{1} \and
Roberto Verdecchia\inst{1}\thanks{Corresponding author.} \and
Enrico Vicario\inst{1}}
\authorrunning{M. Becattini et al.}

\institute{Department of Information Engineering, University of Florence, Florence, Italy\\
\email{marco.becattini@unifi.it}\\
\email{niccolo.caselli3@stud.unifi.it}\\
\email{matteo.minin@stud.unifi.it}\\
\email{roberto.verdecchia@unifi.it}\\
\email{enrico.vicario@unifi.it}}

\maketitle
\begin{abstract}
Automated assessment in software engineering education has advanced significantly for code grading and essay scoring. However, reviewing software architecture deliverables, which requires analyzing structural completeness and requirements traceability, has not yet been fully automated. Applying Large Language Models (LLMs) to this task requires robust architectures to ensure technical feedback is accurate and reliable for students. This paper presents CAPRA (Configurable Architecture Proficiency Report Assessment), a multi-agent LLM system that analyzes software architecture deliverables to generate personalized, template-compliant \LaTeX~feedback. As a core design choice, CAPRA coordinates {multiple specialized agents} and employs a Python-based microservice for multi-modal document extraction, utilizing PyMuPDF and vision-enabled LLMs (specifically \texttt{gpt-4o}) to parse text and UML diagrams. To ensure educational reliability and mitigate hallucinations, CAPRA introduces a deterministic Evidence Anchoring step using fuzzy matching via normalized Levenshtein distance, along with a \texttt{ConsistencyManager} agent that cross-verifies, deduplicates, and merges findings. System performance is assessed using a structured eight-criterion binary evaluation taxonomy covering: (i) extraction completeness, (ii) {feature validation}, (iii) issue grounding and severity detection, (iv) recommendation specificity and traceability, and (v) template and tone compliance. {A preliminary empirical evaluation on 10 student reports shows that CAPRA satisfied 88.8\% of the evaluated criteria under a strict two-rater aggregation rule, achieved moderate inter-rater agreement with human evaluators ($\kappa = 0.582$), and processed each report in {slightly over 4 minutes}.} While these results support the viability of LLM-supported architectural feedback, human oversight remains essential for subjective assessment dimensions.
\end{abstract}

\section{Introduction}

Software engineering education heavily relies on project-based learning, in which students are expected to produce comprehensive documentation, such as requirements specifications, UML diagrams, and architectural designs~\cite{petersen2023capstone}. Reviewing these technically dense artifacts is highly time-consuming, requiring instructors to possess both deep domain knowledge and the capacity to provide tailored, constructive feedback~\cite{hattie2007feedback}. As class sizes grow, this review bottleneck significantly limits the frequency and quality of the formative feedback students receive~\cite{messer2024systematic}, while industry continues to report significant gaps between graduate abilities and workplace expectations in software engineering contexts~\cite{radermacher2013gaps}.

While automated tools have matured significantly for code grading (e.g., through static analysis and unit testing)~\cite{keuning2019systematic,alamutka2005survey}, providing formative feedback on open-ended software architecture documentation remains a difficult challenge. Recent advances in Large Language Models (LLMs)~\cite{brown2020language} have demonstrated promising capabilities for automated review tasks~\cite{xu2025systematic,kasneci2023chatgpt}. However, applying LLMs to complex, multi-modal architectural documents requires careful orchestration to ensure that the generated feedback is reliable, accurate, and pedagogically valuable, avoiding the risk of hallucinated~\cite{ji2023hallucination} or redundant critiques that could misguide students.

{To address these challenges, this paper introduces the Configurable Architecture Proficiency Report Assessment (CAPRA), an automated tool designed to support software engineering students by providing reliable, formative feedback on architectural deliverables without assigning grades. CAPRA adopts a multi-agent architecture in which specialized components handle multi-modal extraction and dimension-specific analysis, while a verification and consistency stage checks cited evidence and consolidates overlapping feedback.}

{Our preliminary empirical evaluation on 10 student reports suggests the viability of the approach. CAPRA satisfied 88.8\% of the evaluated criteria under a strict two-rater aggregation rule. The system processes a complete architectural report in {slightly over 4 minutes} at a cost of roughly \$0.44, compared to an estimated 30--45 minutes for a thorough manual review.}

This contribution is intended for software engineering educators and researchers seeking pedagogical insights into solutions able to provide scalable, high-quality, customized feedback. The main contributions of this work are summarized as follows:
\begin{itemize}
    \item {\textbf{A Multi-Agent Assessment Workflow}: A workflow for software architecture deliverable assessment that integrates multi-modal extraction, dimension-specific analysis, and template-compliant feedback generation.}
    \item {\textbf{Deterministic Evidence Verification}: A mechanism based on fuzzy matching~\cite{approxmatching2001} via normalized Levenshtein distance~\cite{normalizedlevenshtein2007} that checks cited findings against the source document before they are included in the final feedback.}
    \item {\textbf{An Exploratory Empirical Study}: A preliminary evaluation of CAPRA on 10 student project reports, using a custom eight-criterion taxonomy and a separate set of 10 reports for knowledge-base construction.}
\end{itemize}

\section{Related Work}
In this section, we discuss the work more closely related to our research. An overview of the most representative related literature is presented in Table~\ref{tab:relwork} and is further discussed below.

\begin{table}[hbpt]
\centering
\revD{
\caption{Comparison of representative AI/LLM-based assessment and multi-agent approaches across the dimensions relevant to architectural-deliverable review. CAPRA is listed last to position the gap it fills.}
\label{tab:relwork}
\scriptsize
\setlength{\tabcolsep}{2.8pt}
\renewcommand{\arraystretch}{1.18}
\begin{tabular}{>{\raggedright\arraybackslash}m{2.25cm} >{\raggedright\arraybackslash}m{1.55cm} >{\raggedright\arraybackslash}m{2.15cm} >{\raggedright\arraybackslash}m{2.9cm} >{\raggedright\arraybackslash}m{1.65cm}}
\toprule
\textbf{Approach} & \textbf{Artifact} & \textbf{Method} & \textbf{Grounding} & \textbf{Output} \\
\midrule
Static-analysis graders, e.g., ArTEMiS~\cite{krusche2018artemis} & Code & Tests / rules & Deterministic oracle & Grade + feedback \\
LLM UML grading~\cite{bouali2025grading} & UML diagrams & LLM scoring & Compared with TA scores; no source-span anchoring & Score \\
LLM code feedback~\cite{phung2023gpt4} & Code errors & LLM repair + explanation & Runtime / exact-match validation & Feedback \\
LLM short-answer grading~\cite{xie2024grade} & Text answers & Multi-step LLM grading & Rubric + audit loop & Grade \\
LLM-as-a-judge methods~\cite{zheng2023judging,liu2023geval,kim2024prometheus} & Open-ended text & Rubric/reference judging & Human/rubric alignment; no source-span anchoring & Score / preference \\
Generative MAS for software/\allowbreak tasks~\cite{hong2024metagpt,qian2024chatdev,wu2023autogen} & Software tasks & Role-based agents & Internal review / tool checks & Software artifacts \\
\midrule
\textbf{CAPRA (our contribution)} & Architecture reports & Multi-agent assessment & \textbf{Deterministic evidence anchoring} & \textbf{Formative feedback} \\
\bottomrule
\end{tabular}
}
\end{table}

\subsection{Automated Assessment in SE Education}
Automated assessment is widely used to scale programming education: tools like ArTEMiS~\cite{krusche2018artemis} provide immediate feedback through static analysis and unit testing~\cite{ihantola2010review,singh2013automated}, increasingly aided by LLM-based assistants. \revD{Such tools, however, target \emph{deterministic} artifacts. Source code can be checked against an objective oracle (compilation, unit tests), whereas open-ended documentation such as requirements, designs, and UML diagrams cannot, and its assessment remains a persistent challenge}~\cite{messer2024systematic,fan2023llmse}. 

\revD{To bridge this gap, a growing body of work applies Large Language Models~(LLMs) to open-ended artifacts}~\cite{chu2025agents}\revD{, grading UML diagrams}~\cite{bouali2025grading}\revD{, generating feedback on programming syntax errors}~\cite{phung2023gpt4}\revD{, and scoring short-answer responses}~\cite{xie2024grade}\revD{. As Table~\ref{tab:relwork} shows, these efforts generally focus on a single artifact type or representation and do not provide deterministic evidence anchoring of feedback claims to source spans in multi-modal architecture documents. This can produce \emph{hallucinations}, that is, fabricated errors absent from the source}~\cite{xu2025systematic,huang2023hallucination,ji2023hallucination}\revD{, which is especially problematic in education, where incorrect feedback can misguide students}~\cite{hattie2007feedback}.

\subsection{Multi-Agent Systems in SE}
Multi-Agent Systems (MAS) improve performance on complex reasoning tasks~\cite{guo2024multiagent}: frameworks such as MetaGPT~\cite{hong2024metagpt}, AutoGen~\cite{wu2023autogen}, ChatDev~\cite{qian2024chatdev}, CAMEL~\cite{li2023camel}, and AgentVerse~\cite{chen2024agentverse} decompose problems into specialized roles that \revD{can outperform single-agent or monolithic prompting on complex tasks}~\cite{li2024multiagent}. 

\revC{Assessment is a recent MAS application. Grade-Like-a-Human}~\cite{xie2024grade}~{pairs a multi-step grading pipeline with a post-grading audit loop, but grades short-answer text and neither handles multi-modal documents nor anchors findings to verifiable source spans.}

\subsection{LLM-as-a-Judge for Open-Ended Tasks}
The ``LLM-as-a-Judge'' paradigm establishes that frontier LLMs can effectively evaluate open-ended textual outputs~\cite{zheng2023judging}, yet they exhibit systematic biases~\cite{wang2024faireval}, favoring longer texts and being sensitive to information order. Researchers counter this by decomposing it into independent, rubric-driven steps, improving reliability~\cite{xie2024grade,liu2023geval,kim2024prometheus}. \revC{This motivates two design principles for trustworthy assessment: decomposing the evaluation into focused, single-purpose checks, and grounding each {judgment} in concrete evidence rather than an opaque verdict.} 

\section{The CAPRA Approach}

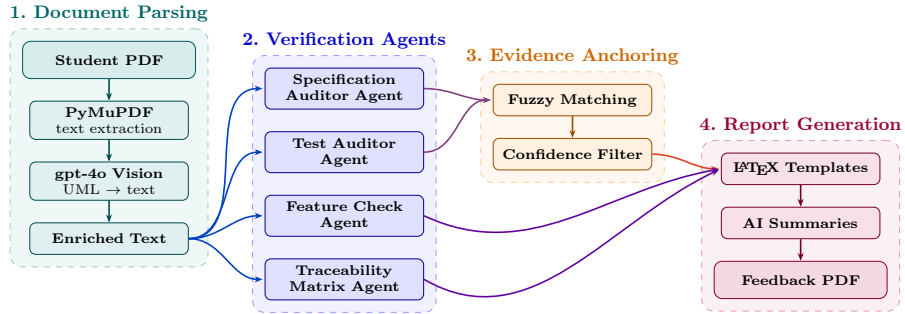
\begin{figure*}[ht]
\centering
\tikzset{
  stageBox/.style={rectangle, rounded corners=4pt, draw=#1!70!black, fill=#1!15,
                   text width=2.8cm, minimum height=0.65cm, align=center,
                   font=\scriptsize\bfseries},
  agentBox/.style={rectangle, rounded corners=3pt, draw=#1!60!black, fill=#1!12,
                   text width=2.55cm, minimum height=0.55cm, align=center,
                   font=\scriptsize},
  stageLabel/.style={font=\footnotesize\bfseries, text=#1!80!black},
  myArrow/.style={-{Stealth[length=4pt]}, thick, #1},
  stageBg/.style={rectangle, rounded corners=6pt, draw=#1!40, fill=#1!6,
                  inner sep=6pt, dashed},
}
\resizebox{0.98\textwidth}{!}{%
\begin{tikzpicture}[node distance=0.38cm and 0.55cm]

\node[stageBox=teal] (pdf) {Student PDF};
\node[agentBox=teal, below=of pdf]
  (pymupdf)  {\textbf{PyMuPDF}\\\scriptsize{text extraction}};
\node[agentBox=teal, below=of pymupdf]
  (vision)   {\textbf{gpt-4o Vision}\\\scriptsize{UML $\to$ text}};
\node[agentBox=teal, below=of vision]
  (enriched) {\textbf{Enriched Text}};

\begin{scope}[on background layer]
  \node[stageBg=teal, fit=(pdf)(enriched),
        label={[stageLabel=teal]above:\textbf{1. Document Parsing}}]
        (stage1) {};
\end{scope}

\node[agentBox=blue, right=1.2cm of pdf, yshift=-0.5cm]
  (specAgt)  {\textbf{Specification}\\\textbf{Auditor Agent}};
\node[agentBox=blue, below=of specAgt]
  (testAgt)  {\textbf{Test Auditor}\\\textbf{Agent}};
\node[agentBox=blue, below=of testAgt]
  (featAgt)  {\textbf{Feature Check}\\\textbf{Agent}};
\node[agentBox=blue, below=of featAgt]
  (traceAgt) {\textbf{Traceability}\\\textbf{Matrix Agent}};

\begin{scope}[on background layer]
  \node[stageBg=blue, fit=(specAgt)(traceAgt),
        label={[stageLabel=blue]above:\textbf{2. Verification Agents}}]
        (stage2) {};
\end{scope}

\node[agentBox=orange, right=1.2cm of specAgt, yshift=-0.2cm]
  (fuzzy)  {\textbf{Fuzzy Matching}};
\node[agentBox=orange, below=of fuzzy]
  (filter) {\textbf{Confidence Filter}};

\begin{scope}[on background layer]
  \node[stageBg=orange, fit=(fuzzy)(filter),
        label={[stageLabel=orange]above:\textbf{3. Evidence Anchoring}}]
        (stage3) {};
\end{scope}

\node[agentBox=purple, right=1.2cm of fuzzy, yshift=-1.2cm]
  (latex)  {\textbf{\LaTeX{} Templates}};
\node[agentBox=purple, below=of latex]
  (aiSumm) {\textbf{AI Summaries}};
\node[stageBox=purple, below=of aiSumm]
  (pdfout) {Feedback PDF};

\begin{scope}[on background layer]
  \node[stageBg=purple, fit=(latex)(pdfout),
        label={[stageLabel=purple]above:\textbf{4. Report Generation}}]
        (stage4) {};
\end{scope}

\draw[myArrow=teal!60!black]   (pdf)    -- (pymupdf);
\draw[myArrow=teal!60!black]   (pymupdf)-- (vision);
\draw[myArrow=teal!60!black]   (vision) -- (enriched);
\draw[myArrow=orange!60!black] (fuzzy)  -- (filter);
\draw[myArrow=purple!60!black] (latex)  -- (aiSumm);
\draw[myArrow=purple!60!black] (aiSumm) -- (pdfout);

\foreach \agent in {specAgt, testAgt, featAgt, traceAgt}{
  \draw[myArrow=teal!50!blue, shorten >=1pt]
    (enriched.east) to[out=0,in=180] (\agent.west);
}
\draw[myArrow=blue!50!orange, shorten >=1pt]
  (specAgt.east)  to[out=0,in=180] (fuzzy.west);
\draw[myArrow=blue!50!orange, shorten >=1pt]
  (testAgt.east)  to[out=0,in=180] (fuzzy.west);
\draw[myArrow=orange!50!purple, shorten >=1pt]
  (filter.east)   to[out=0,in=180]  (latex.west);
\draw[myArrow=blue!50!purple, shorten >=1pt]
  (featAgt.east)  to[out=-25,in=190] (latex.west);
\draw[myArrow=blue!50!purple, shorten >=1pt]
  (traceAgt.east) to[out=-30,in=205] (latex.west);

\end{tikzpicture}%
}
\caption{CAPRA system architecture: four-stage pipeline from PDF ingestion
(Document Parsing), through parallel multi-agent evaluation
(Verification Agents), evidence-anchored deduplication
(Evidence Anchoring), to final report generation (Report Generation).}
\label{fig:architecture_v4}
\end{figure*}

This section details the architecture and operational workflow of CAPRA. To address the complexity of educational software architecture assessment, the system is designed as a pipeline divided into four main stages (see Figure~\ref{fig:architecture_v4}): Document Parsing, Parallel Verification, Evidence Anchoring, and Report Generation.

\subsection{Document Parsing and Extraction}
\textbf{Input:} Student's raw architectural deliverable (PDF format).\newline
\textbf{Output:} Enriched text representation of the document.

This phase takes the student's PDF report and converts it into a single text format. First, it extracts the main text using the PyMuPDF library \cite{pymupdf2024}. Since architectural documents contain many visual elements like UML diagrams, standard text extraction is not enough~\cite{dotsocr2025}. To solve this, the system uses the \texttt{gpt-4o} vision model~\cite{openai2024gpt4o} to parse images and turn them into structured text descriptions. Finally, these descriptions are inserted into the original text exactly where the images appeared. This creates an enriched, sequential text document that preserves the original flow, making it ready for the next analysis steps.

\subsection{Parallel Verification Agents}
\textbf{Input:} Enriched text representation of the document.\newline
\textbf{Output:} A collection of raw analytical findings, each with textual evidence and a confidence score.\newline
This stage takes the enriched text as input and analyzes it to identify architectural issues. A central orchestrator activates several specialized AI agents that run at the same time. Each agent examines the entire document, but focuses only on a small task~\cite{hong2024metagpt,li2024multiagent}. This strategy provides a deep analysis without the confusion that often happens when a single prompt tries to do everything at once.

\begin{figure}[ht]
\centering
\tikzset{
  stageBox/.style={rectangle, rounded corners=4pt, draw=#1!70!black, fill=#1!15,
                   text width=2.8cm, minimum height=0.65cm, align=center,
                   font=\scriptsize\bfseries},
  agentBox/.style={rectangle, rounded corners=3pt, draw=#1!60!black, fill=#1!12,
                   text width=2.55cm, minimum height=0.55cm, align=center,
                   font=\scriptsize},
  stageLabel/.style={font=\footnotesize\bfseries, text=#1!80!black},
  myArrow/.style={-{Stealth[length=4pt]}, thick, #1},
  stageBg/.style={rectangle, rounded corners=6pt, draw=#1!40, fill=#1!6,
                  inner sep=6pt, dashed},
}
\resizebox{0.88\textwidth}{!}{%
\begin{tikzpicture}[node distance=0.38cm and 0.55cm]

\node[agentBox=teal] (f1) {``UI mockups for interfaces''};
\node[agentBox=teal, below=of f1] (f2) {``Navigation menu design''};
\node[agentBox=teal, below=of f2] (f3) {``Input validation forms''};
\node[font=\tiny\itshape, text=gray, below=0.2cm of f3] (simLabel)
     {\textit{n} similar features};

\begin{scope}[on background layer]
  \node[stageBg=teal, fit=(f1)(f2)(f3)(simLabel),
        label={[stageLabel=teal]above:\textbf{Feature Cluster}}]
        (clusterBox) {};
\end{scope}

\node[stageBox=orange, right=1.4cm of clusterBox] (llm)
     {LLM Compactor\\\scriptsize\textit{merge \& synthesize}};

\begin{scope}[on background layer]
  \node[stageBg=orange, fit=(llm),
        label={[stageLabel=orange]above:\textbf{Compaction}}]
        (compactBox) {};
\end{scope}

\node[agentBox=purple, right=1.4cm of compactBox, text width=3.8cm,
      align=left, inner sep=5pt] (summContent)
{
  \textbf{feature:} ``UI \& Interaction Design''\\[2pt]
  \textbf{desc:} ``Contains UI mockups and clear navigation structures\ldots''\\[2pt]
  \textbf{checklist:}\\[1pt]
  \hspace{3pt}$\checkmark$ UI mockups for key interfaces\\[0.5pt]
  \hspace{3pt}$\checkmark$ Clear navigation elements\\[0.5pt]
  \hspace{3pt}$\checkmark$ Input validation mechanisms\\[0.5pt]
  \hspace{3pt}$\checkmark$ Structured input fields\\[0.5pt]
  \hspace{3pt}\textcolor{gray}{\ldots}
};

\begin{scope}[on background layer]
  \node[stageBg=purple, fit=(summContent),
        label={[stageLabel=purple]above:\textbf{Summary Feature}}]
        (summBox) {};
\end{scope}

\foreach \feat in {f1, f2, f3}{
  \draw[myArrow=teal!50!orange, shorten >=1pt]
    (\feat.east) to[out=0,in=180] (llm.west);
}
\draw[myArrow=orange!50!purple, shorten >=1pt]
  (llm.east) to[out=0,in=180] (summContent.west);

\end{tikzpicture}%
}
\caption{LLM Compactor: Similar features within a cluster (e.g., ``UI mockups for interfaces'', ``Navigation menu design'', ``Input validation forms'') are merged by an LLM into a single canonical Summary Feature with a structured checklist used by the \textit{FeatureCheckAgent}.}
\label{fig:compactor}
\end{figure}
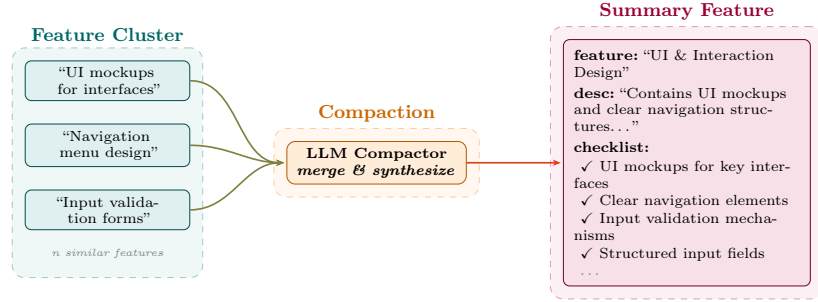

To guarantee reliable and {consistent feedback}, the system forces the language models to respond deterministically (by setting a seed for OpenAI's models and temperature zero for all the models used). When an agent finds a problem, it reports the issue, attaches a direct quote from the document as proof, and assigns a confidence score. The main agents are:
\begin{itemize}
    \item \textit{SpecificationAuditorAgent}: Performs a comprehensive audit to identify flaws in functional and non-functional requirements (e.g., incompleteness, business logic contradictions) and use case specifications (template structure, actor roles). It also verifies the alignment between UML diagrams and the textual narrative, and evaluates the architectural description for pattern correctness and separation of concerns, providing literal quotes as evidence for each finding.
    \item \textit{TestAuditorAgent}: Audits the test plan to identify critical coverage gaps and verify adherence to the stated testing strategy, extracting relevant evidence quotes.
    \item \textit{FeatureCheckAgent}: Checks if the document contains the required features based on a grading rubric stored in a database. Instructors can write these rules manually, or the system can automatically extract them from past student reports. In our project, we used a SALLMA workflow~\cite{sallma2025} to extract all the features of historical documents. The system then groups similar features together using a clustering algorithm, namely HDBSCAN~\cite{campello2013hdbscan}, and keeps only the clusters that are present in at least two-thirds of the reports. For each valid cluster, a representative feature is selected to generate a specific checklist ({see Figure~\ref{fig:compactor}}), which is then used to verify if that feature is present in a new report. This allows instructors to easily update the grading rules as the course evolves.
    \item \textit{Traceability Matrix Agent}: Analyzes the full document text to build a comprehensive mapping matrix (\texttt{Requirement} $\rightarrow$ \texttt{UseCase} $\rightarrow$ \texttt{Design} $\rightarrow$ \texttt{Test}) and identify architectural disconnects.
\end{itemize}

\subsection{Evidence Anchoring and Deduplication} \label{sec:evidence_anchoring}
\textbf{Input:} The collection of raw analytical findings.\newline
\textbf{Output:} A verified, evidence-backed set of architectural critiques.\newline
The \textit{Evidence Anchoring Service} is a deterministic algorithm that checks if the student actually wrote what the agent claims.

This stage connects the raw issues found by the agents with the actual proof in the report. Specifically, it takes the findings of the two main review agents (\textit{SpecificationAuditorAgent} and \textit{TestAuditorAgent}), each carrying an initial confidence score ($C_{initial}$) assigned by the originating agent, and modulates that score based on how well the agent's cited quote matches the original document text.

After cleaning up the text by removing punctuation and extra spaces, each issue is handled based on the citation length as follows:
\begin{itemize}
    \item \textbf{Missing or Empty Quote:} If there is no quote, the issue is not immediately discarded, but its confidence score is halved (50\% penalty). This conservative choice avoids losing potentially valid issues that the agent failed to ground properly.
    \item \textbf{Short Quote ($\mathbf{<}$15 characters):} Very short texts are hard to match safely. Instead of running a complex search, the system applies a 30\% penalty to the confidence score (multiplying it by 0.7). This keeps good findings alive without trusting them blindly.
    \item \textbf{Standard Quote:} The system scans the document using a sliding window ($W$) of length $|Q|$ to find the quote ($Q$). It attempts an exact match first. If that fails, it applies a fast trigram overlap pre-filter. A trigram is a sequence of 3 characters. The system extracts all trigrams from $W$ and counts how many of $Q$'s trigrams are present within them. This number is divided by the total number of trigrams in $Q$, producing an overlap ratio from 0 to 1. Windows with an overlap below 0.27 are immediately discarded. Because calculating the precise Levenshtein distance~\cite{levenshtein1966binary} is computationally expensive, this fast trigram check efficiently filters out completely different paragraphs, reserving the true similarity calculation only for highly probable matches.
\end{itemize}

For a given extracted window $W$ and the LLM-generated quote $Q$, the normalized similarity is mathematically defined as~\cite{marzal1993computation}:
\begin{equation}
Sim(W, Q) = 1 - \frac{Levenshtein(W, Q)}{\max(|W|, |Q|)}
\end{equation}

\begin{figure}[ht]
\centering
\includegraphics[width=0.76\columnwidth]{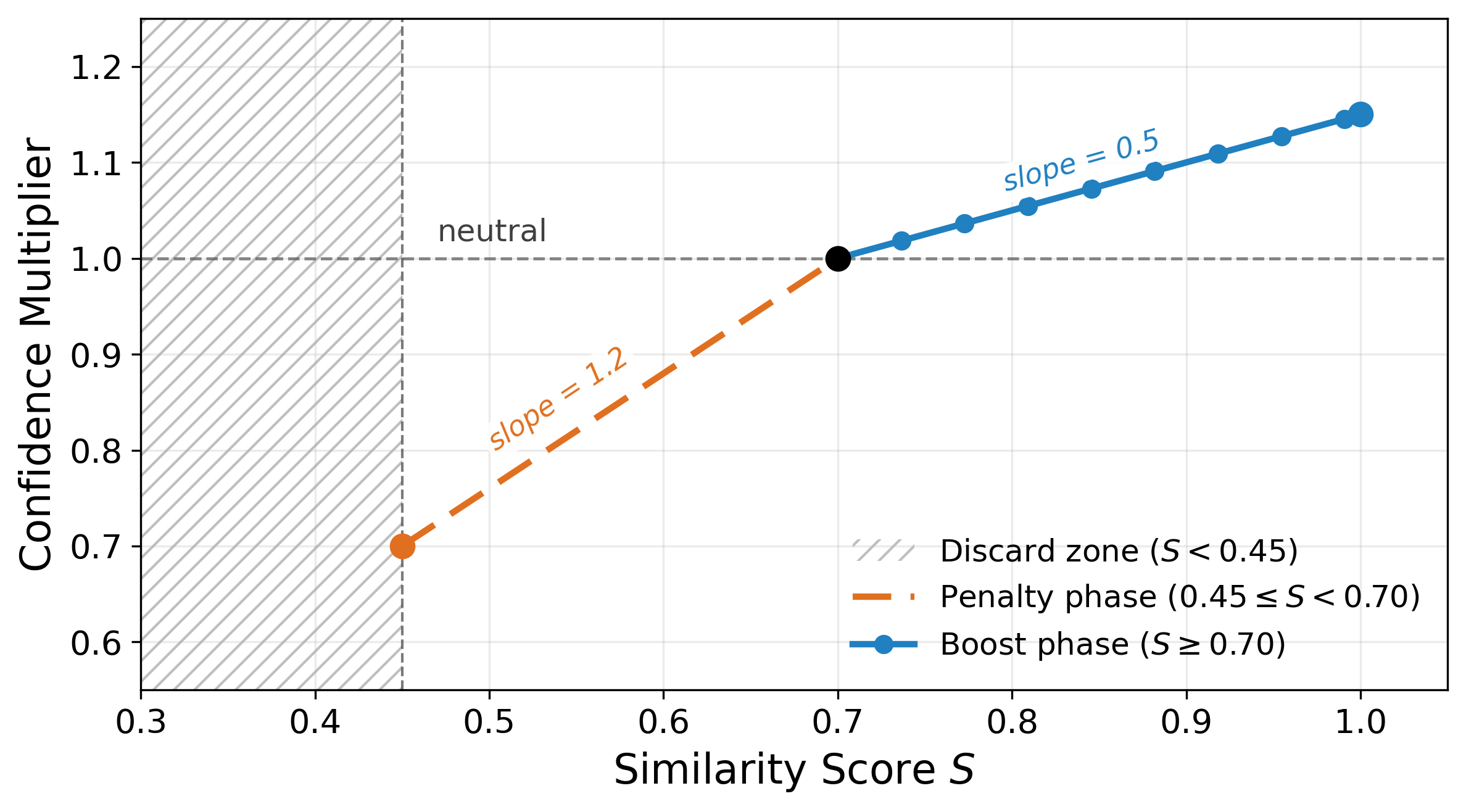}
\caption{Confidence modulation function: the multiplier applied to the initial confidence score $C_i$ as a function of the similarity score $S$.}
\label{fig:confidence_modulation}
\end{figure}

For each candidate issue, the system verifies its textual grounding using the maximum similarity score $S = Sim_{max}$ calculated across all candidate document windows. This grounding is processed through a two-tier verification logic:
\begin{enumerate}
    \item \textbf{Discard Threshold ($\tau_{min} = 0.45$):} If $S < \tau_{min}$, the quote is considered non-verifiable (a potential hallucination) and the entire issue is discarded.
    \item \textbf{Confidence Modulation:} For all surviving issues ($S \ge 0.45$), the initial confidence score $C_i$ is updated to a modulated value $C_{new}$ {based on the similarity score, adjusting down for poor matches and slightly boosting for high matches (see Figure~\ref{fig:confidence_modulation}).}

    {To illustrate the effect of similarity on the final confidence score, consider a few examples. If the similarity is very high ($S \ge 0.70$), the confidence score slightly increases. For example, with an initial confidence of 0.85, a perfect match ($S = 1.0$) boosts it to $\sim 0.98$. If similarity drops to 0.70, it stays at 0.85. Below 0.70, a penalty logic is activated. For instance, with an initial confidence of 0.85, a similarity of 0.50 reduces the confidence score to $\sim 0.65$.}
\end{enumerate}

After adjusting the scores, a strict {\textit{Confidence Filter}} deletes any issue that scores below 0.65. The remaining solid findings go to the {\textit{ConsistencyManager}}, inspired by the self-consistency principle of cross-checking multiple outputs to improve reliability~\cite{wang2023selfconsistency}. This is a final AI agent that reads all the critiques, merges repeated ones (e.g. combining multiple repeated issues into a single clear issue), and gives the final approval. Issues are then sorted by severity and renumbered sequentially by category, keeping the feedback focused and useful.

\subsection{Report Generation}
\textbf{Input:} Verified architectural critiques and extracted document entities.\newline
\textbf{Output:} A formative pedagogical feedback report in PDF format.\newline
The final phase transforms the verified issues into a well-formatted pedagogical PDF report. By relying on deterministic \LaTeX~templates rather than generating the entire document via LLM, CAPRA ensures that the compilation process is significantly faster, cheaper (due to drastically reduced token usage), and free of formatting errors. The system combines these pre-built structural skeletons with three targeted AI requests, which are delegated to a lightweight, cost-efficient model (\texttt{claude-haiku-4.5}). Since these requests produce short natural-language narratives that are injected into pre-validated \LaTeX~skeletons, rather than raw \LaTeX~code which remains a challenging task for many LLMs, a smaller model is sufficient for this stage. These requests write short narrative summaries:
\begin{enumerate}
    \item \textbf{Document Context:} Synthesizes a structured overview of the system under analysis, covering its objectives, main use cases, functional and non-functional requirements, architecture, and testing strategy.
    \item \textbf{Executive Summary:} Provides a concise narrative framing the overall quality of the document, highlighting the patterns of issues found, the critical areas, and the top priority actions for the student.
    \item \textbf{Strengths:} Identifies the positive aspects of the student's work (e.g., clear structure, good use case coverage, thorough testing), each grounded in concrete evidence from the document.
\end{enumerate}

{All other sections, such as the feature coverage checklist, issue details, traceability matrix, and the audit findings themselves, are populated deterministically using the structured data extracted by the agents.} Finally, a compilation service renders the \texttt{pdflatex} document, delivering a structured and actionable feedback report to the student.

\section{Exploratory Empirical Study Design}

To assess the viability of CAPRA as a teaching assistant, we designed our preliminary evaluation to answer the following three Research Questions (RQs):
\begin{itemize}
    \item {\textbf{RQ1 (Feedback Quality and Reliability):} \textit{How valid, reliable, and constructive is the feedback generated by CAPRA?}}
    \item \textbf{RQ2 (Processing Efficiency):} \textit{How does CAPRA's processing time compare to the manual review effort typically required for software architecture deliverables?}
    \item {\textbf{RQ3 (System Customization):} \textit{How well does CAPRA identify the presence or absence of Knowledge Base features within student architectural deliverables?}}
\end{itemize}

\subsection{Study Design and Data Collection}
To rigorously answer the RQs, our empirical evaluation was structured into three distinct phases, following established guidelines for empirical research in software engineering~\cite{kitchenham2002guidelines,runeson2009guidelines}.

\smallskip\noindent\textbf{Phase 1: Experimental Context}\quad The study examines project reports from the Software Engineering course within the
Bachelor of Science (BSc) Computer Engineering program at the University of Florence, 
a course that has been running for more than 20 years (more than 5 years in its 
current form), and typically enrolls approximately 75 students per academic year. 
We selected a sample of 10 student reports, produced in recent academic years and 
awarded the highest grades in the course, which served to construct the knowledge 
base (Figure~\ref{fig:compactor}). High-scoring reports were deliberately chosen 
for this role as they represent well-structured, complete deliverables, allowing 
the system to extract high-quality, representative features without noise introduced 
by structural deficiencies. A second, separate sample of 10 student reports was then 
employed for evaluation purposes. Deliverable documents of the considered course are 
notably complex and lengthy, {typically ranging between 10 and 70 pages.}

\smallskip\noindent\textbf{Phase 2: Data Collection Framework}\quad To systematically assess the output of the CAPRA pipeline, we defined a comprehensive rubric consisting of eight distinct evaluation criteria {organized} into four dimensions, as detailed in Table~\ref{tab:taxonomy}. These criteria investigated both the extraction capabilities and the qualitative assessment traits of the multi-agent system.

\begin{table}[!htbp]
\caption{Evaluation taxonomy: dimensions, criteria IDs, and binary questions used by the two independent raters (Pass~=~1, Fail~=~0).}
\centering
\scriptsize
\begin{tabular*}{\textwidth}{@{\extracolsep{\fill}} >{\centering\arraybackslash}m{2.35cm} >{\centering\arraybackslash}m{0.7cm} >{\raggedright\arraybackslash}m{8.15cm} @{}}
\toprule
\textbf{Dim.} & \textbf{ID} & \textbf{Binary evaluation question} \\
\midrule
\multirow{4}{*}{\shortstack{A\\Extraction\\Completeness}}
  & A1 & Were all functional \& non-functional requirements correctly identified and extracted? \\
  & A2 & Were all use cases extracted with actors, pre/post-conditions, and alternative flows? \\
  & A3 & Were all architectural components, layers, and design patterns correctly identified? \\
  & A4 & Were all test types (unit/integration) and coverage metrics correctly extracted? \\
\midrule
\shortstack{B\\Feature\\Validation}
  & B1 & Are the identified Software Engineering (SWE) features real and present, and is each checklist status (Present/Partial/Absent) correctly assigned? \\
\midrule
\multirow{2}{*}{\shortstack{C\\Issues \& Recomm.}}
  & C1 & Are reported issues real, grounded in source evidence, with no false positives and no missed HIGH-severity problems? \\
  & C2 & Are recommendations specific and directly tied to each identified issue? \\
\midrule
\shortstack{D\\Template\\\& Tone}
  & D1 & Are all expected sections present and written in formal, objective tone? \\
\bottomrule
\end{tabular*}
\label{tab:taxonomy}
\end{table}

\smallskip\noindent\textbf{Phase 3: Data Analysis Strategy}\quad The feedback generated by CAPRA for each of the 10 reports was independently reviewed by two authors of this article. Both raters evaluated the tool's output for each category in a binary manner (1 for Pass/Valid, 0 for Fail/Invalid). {We adopted binary ratings in this exploratory study to simplify cross-rater comparison and aggregation across criteria, and to keep the evaluation protocol reproducible and easy to calibrate across evaluators. These human ratings serve as the reference against which CAPRA's output quality is interpreted; consequently, inter-rater agreement quantifies the reliability of that reference and provides context for judging how meaningful CAPRA's observed agreement levels are.} To scientifically measure the consistency of their judgments, we computed Cohen's Kappa coefficient ($\kappa$)~\cite{cohen1960kappa}. This metric accounts for chance agreement, providing a more reliable measure of consistency than a simple percentage~\cite{landis1977kappa}.

\section{Results}
\label{results}

In this section, we present the objective findings from our empirical evaluation, structured according to the research questions.

\subsection{Accuracy and Reliability of Architectural Feedback}
\label{sec:rq1_results}
Overall, CAPRA produced structurally coherent, template-compliant feedback reports for all submissions. The system was particularly strong on \emph{extractive} tasks (e.g., requirements and use-case identification) and on surfacing testing-related anomalies in a consistent manner. Importantly, the Evidence Anchoring mechanism acted as a {verification gate}: findings that could not be verified against the source text were filtered out, substantially reducing ungrounded critiques.

\textbf{Pass rates under lenient vs.\ strict aggregation.}
Table~\ref{tab:results} reports two complementary pass-rate aggregations computed from the two human raters' binary {judgments} on CAPRA’s output.
{For a given report and criterion, let $r_1, r_2 \in \{0,1\}$ denote the two binary rater scores. The lenient aggregation is computed as $\mathrm{Avg.} = \frac{r_1 + r_2}{2}$, whereas the strict aggregation is defined as $\mathrm{Min.} = \min(r_1, r_2)$. Thus, \textit{Avg.} can take values $\{0, 0.5, 1\}$, while \textit{Min.} equals 1 only when both raters assign \textit{Pass}.} Under strict aggregation, the overall pass rate is \textbf{88.8\%}; under lenient aggregation it is \textbf{91.9\%}.

\textbf{Inter-rater reliability and class imbalance.}
Across all 80 aligned rater decisions (8 categories $\times$ 10 reports), the two raters reached unanimous agreement on 75 cases, yielding a raw agreement rate of \textbf{93.75\%}. However, this figure must be interpreted with caution: the evaluation corpus consists exclusively of high-scoring reports, so the majority of decisions are \textit{Pass} across almost every category. This strong class imbalance inflates raw agreement, since both raters are likely to agree on the dominant class regardless of any genuine alignment in {judgment}. Cohen's Kappa corrects for this chance agreement and is therefore the primary reliability metric we report.

Where the two diverge (A3, C1, D1), genuine inter-rater disagreements occurred, directly mirrored by the $\kappa$ values. For four categories (A2, A4, B1, C2) all 10 decisions were unanimous \textit{Pass}: $\kappa$ is undefined due to zero variance (\textit{n/a}). The global $\kappa = 0.582$ places CAPRA at the upper bound of \textit{moderate agreement}~\cite{landis1977kappa} once class imbalance is accounted for.

These figures highlight how different architectural aspects behave under automated evaluation. For extractive tasks like \textit{Requirements Extraction} (A1, $\kappa = 1.00$), both raters consistently agreed, confirming that structured information retrieval is a clear strength of the pipeline. Architecture \& Design Patterns (A3, $\kappa = 0.615$) showed moderate agreement, likely reflecting the inherently interpretive nature of assessing architectural descriptions against a rubric. \textit{Grounded Issues} (C1, $\kappa = 0.348$, \textit{fair}) is the most structurally complex criterion: it simultaneously tests for false positives (fabricated issues) and false negatives (missed HIGH-severity problems), two failure modes that individual raters weighted differently. \textit{Template \& Tone} (D1, $\kappa = 0.000$) is a well-known statistical {artifact}: with the first reviewer assigning 9 passes and the second reviewer 10, the single disagreement falls in precisely the configuration where expected and observed agreement coincide, collapsing $\kappa$ to zero despite 90\% raw agreement. This interaction between class imbalance and Kappa is a known limitation of the metric~\cite{landis1977kappa} and underscores the importance of reporting raw agreement alongside $\kappa$.

\begin{table}[!ht]
\caption{Consolidated results: pass rates under {lenient (\textbf{Avg.}, arithmetic mean of the two binary ratings)} and {strict (\textbf{Min.}, minimum of the two ratings)} aggregation, per-criterion raw agreement, and Cohen's $\kappa$. \textit{n/a}: zero variance (all decisions unanimous \textit{Pass}); $\kappa$ undefined. See Table~\ref{tab:taxonomy} for criterion definitions.}
\centering
\footnotesize
\begin{tabular*}{\textwidth}{@{\extracolsep{\fill}} c l c c c c @{}}
\toprule
\textbf{ID} & \textbf{Criterion} & \textbf{Avg.\ (\%)} & \textbf{Min (\%)} & \textbf{Raw agr.} & \textbf{Cohen's $\kappa$} \\
\midrule
A1 & Req.\ Extraction Completeness  & 90  & 90  & 100\% & 1.000 \\
A2 & Use Case Completeness           & 100 & 100 & 100\% & \textit{n/a} \\
A3 & Architecture \& Design Patterns & 85  & 80  &  90\% & 0.615 \\
A4 & Test Types \& Coverage Metrics  & 100 & 100 & 100\% & \textit{n/a} \\
B1 & Feature Validation              & 100 & 100 & 100\% & \textit{n/a} \\
C1 & Grounded Issues                 &  65 &  50 &  70\% & 0.348 \\
C2 & Actionable Recommendations      & 100 & 100 & 100\% & \textit{n/a} \\
D1 & Formal Tone \& Template         &  95 &  90 &  90\% & 0.000 \\
\midrule
\textbf{Overall} & & \textbf{91.9} & \textbf{88.8} & \textbf{93.75\%} & \textbf{0.582} \\
\bottomrule
\end{tabular*}
\label{tab:results}
\end{table}

\smallskip
{\small\textbf{Concrete Example of CAPRA Detecting a Semantic Inconsistency}\par\noindent In a ticketing system specification, a specific use case UC-7 (\textit{Ticket Purchase}) is defined with \textit{Guest} as the actor. Yet the Use Case (UC) template includes an alternative flow condition that only makes sense for staff management: \textit{``If the user is not in the event staff: do nothing.''}. This introduces a semantic contradiction between the declared actor role and the described behavior, strongly suggesting a copy-paste artifact in the requirements documentation. \textbf{Human Rater Action.} While reviewing a long specification with multiple use cases, the human evaluator overlooked that the alternative flow constraint is incompatible with the UC’s actor semantics. \textbf{CAPRA Output:}\par\nobreak\smallskip
{\footnotesize\ttfamily
\begin{tabular}{@{}l@{}}
{ISS-004 - HIGH [100\%] - Page 8}\\
UC-7 (Ticket Purchase) contains an evidently\\
incorrect and contradictory alternative flow:\\
``If the user is not\\
in the event staff: do nothing.'' \ldots appears to be a copy-paste\\
from staff management use cases and conflicts with Actors: Guest.
\end{tabular}}\par\smallskip
\noindent{\small\textbf{Result.} This critique demonstrates CAPRA’s effectiveness at identifying UML/UC-level semantic inconsistencies that undermine requirements correctness, highlighting subtle specification defects that are easy to miss in manual reviews.}}
\smallskip

\smallskip
{\small\textbf{Concrete Example of CAPRA Detecting an Architectural Gap}\par\noindent In a barber shop project, the Domain Model defines an abstract \texttt{User} superclass with \texttt{Customer} and \texttt{Barber} subclasses. However, the persistence design stores both roles in a single relational table: \texttt{Users(email, name, surname, pass\_hash, phone, role)}. The report never documents the mapping rule between the OO inheritance hierarchy and the relational schema (e.g., how the \texttt{role} field is consistently used to instantiate the correct subclass across all DAOs). \textbf{Human Rater Action.} The human evaluator, focusing on the presence of both a UML domain model and a database schema, overlooked that the deliverable does not specify any explicit inheritance-to-table mapping strategy, leaving the cross-layer design rationale implicit and non-verifiable. \textbf{CAPRA Output:}\par\nobreak\smallskip
{\footnotesize\ttfamily
\begin{tabular}{@{}l@{}}
{ISS-001 - LOW [100\%] - Page 36}\\
\ldots Users table stores both roles in a single table\\
without a discriminator-based mapping strategy being described \ldots\\
The document does not clarify how the role field is mapped\\
to the correct subclass in all DAOs \ldots
\end{tabular}}\par\smallskip
\noindent{\small\textbf{Result.} This critique demonstrates CAPRA’s ability to detect cross-layer architectural gaps where UML-level abstractions are declared but not formally connected to persistence design decisions, weakening architectural traceability and verifiability.}}
\smallskip

\smallskip\noindent{\footnotesize\textbf{Answer to RQ1 (Highlights).} CAPRA’s outputs satisfied \textbf{88.8\%} of criteria under strict aggregation (\textbf{91.9\%} under lenient aggregation). Human judges agreed on \textbf{93.75\%} of the 80 evaluation decisions; however, this figure is inflated by the Pass-dominant (high-quality) corpus. After correcting for chance agreement, overall human-assessment reliability is $\kappa = 0.582$ (\textit{moderate}). The most challenging dimension is C1 (\textit{Grounded Issues}), with $\kappa = 0.348$ (\textit{fair}) and a \textbf{65\%} pass rate under lenient aggregation.}\smallskip

\subsection{Processing Efficiency (RQ2)}
\label{sec:rq2_results}
To evaluate CAPRA's practical viability as a teaching assistant, we measured end-to-end processing times across the 10 evaluated reports.

The average total processing time was {slightly over 4 minutes per report} (mean = 248s, SD = 73.1s across the 10 reports). 
Document complexity ({page count ranging from 10 to 70 pages}) had a moderate impact on processing time, with longer reports requiring up to 6 minutes.

In terms of API cost, processing a single report consumed an average of approximately \$0.39 in OpenAI API calls (using \texttt{gpt-5.1} pricing) and \$0.04 in Anthropic API calls (using \texttt{claude-haiku-4.5} pricing), {for a total of roughly \$0.44 on average for each report after rounding.}

{The initial system setup requires minimal manual effort: no explicit model training or fine-tuning is needed. Instructors only provide a set of high-quality reports to automatically generate the evaluation Knowledge Base (KB). In our setup, the automatic extraction of the KB from 10 reference reports required processing roughly 786K input tokens and 130K output tokens, incurring a one-time setup cost of approximately \$0.20.}

Compared to the 30--45 minutes typically required for a thorough manual review by an instructor, CAPRA achieves a speedup factor of approximately 7.2--10.8$\times$. For a class of 30 student groups, this translates from 15--22.5 hours of manual review effort to approximately 2 hours of automated processing, perhaps to be complemented by a concise manual review, enabling instructors to provide high-frequency formative feedback cycles.

\smallskip\noindent{\footnotesize\textbf{Answer to RQ2 (Highlights).} CAPRA processes a complete architectural deliverable in {slightly over 4 minutes} ($\sim$\$0.44 per report), achieving a 7.2--10.8$\times$ speedup over manual review (30--45 min).}\smallskip

\subsection{System Customization (RQ3)}
\label{sec:rq3_results}
To evaluate the configurability of CAPRA, we examined how the Knowledge Base (KB) generation pipeline enables automatic adaptation to course-specific rubrics. As described in Section~3.2, the \textit{FeatureCheckAgent} relies on a checklist mined from historical student reports using a SALLMA~\cite{sallma2025} workflow with HDBSCAN clustering.

From the 10 reference reports used for KB generation, the SALLMA extraction phase identified a total of 680 raw feature mentions across all documents.
After HDBSCAN clustering with a minimum cluster size of 3, {applying the two-thirds prevalence threshold (i.e., features present in at least 7 out of 10 reports) reduced the validated features to 7}, each represented by a canonical description selected from a cluster representative. The resulting checklist is reported in Table~\ref{tab:features}.

\begin{table}[!htbp]
\caption{Validated Features Extracted via Automated KB Generation}
\centering
\footnotesize
\begin{tabular}{l c}
\toprule
\textbf{Feature} & \textbf{Mentions} \\
\midrule
Separation of concerns in architecture & 85 \\
Unit testing framework implementation  & 63 \\
UI and interaction design principles   & 56 \\
Use of UML diagrams for system modeling & 36 \\
Data Access Object (DAO) pattern        & 25 \\
Definition and documentation of use cases & 20 \\
Identification and definition of system actors & 8 \\
\midrule
\textbf{Total} & \textbf{293} \\
\bottomrule
\end{tabular}
\label{tab:features}
\end{table}

The effectiveness of this automatically generated KB was assessed by analyzing the \textit{FeatureCheckAgent}'s performance on our evaluation corpus. As reported in Table~\ref{tab:results}, the agent achieved a {100\% pass rate on category B1 (Feature Validation) under the adopted human evaluation protocol}, with unanimous agreement between the two human raters. {This was determined via manual cross-referencing: raters verified that features flagged as ``present'' existed in the documents, and confirmed that those flagged as ``missing'' were absent from the documents. This confirms the agent correctly evaluated feature presence and absence.}

The KB generation process is fully automated and requires only a set of reference reports as input, with no manual rubric authoring. However, the resulting checklist can also be manually refined by the instructor, for example, to add domain-specific criteria, remove irrelevant features, or adjust the granularity of individual items, allowing a hybrid approach that combines automated extraction with human revision. An instructor can regenerate the KB at the start of each academic year by providing new reference materials, ensuring that CAPRA remains aligned with evolving course requirements.

\smallskip\noindent{\footnotesize\textbf{Answer to RQ3 (Highlights).} CAPRA achieved a {100\% pass rate on feature validation (B1) under the adopted human evaluation protocol}, with unanimous inter-rater agreement. {Across all 10 evaluated reports, both raters judged the presence or absence of all 7 KB-generated features as correctly identified by the system.} The KB was automatically generated from 680 raw mentions via HDBSCAN clustering, requiring no manual rubric authoring.}\smallskip

\section{Discussion}

The results presented in Section~\ref{results} paint a nuanced picture of the capabilities and current limitations of LLM-based multi-agent systems for educational assessment. In this section, we interpret the key findings and discuss their implications.

\textbf{Interpretation of Key Results.}
{Overall, CAPRA generates valid architectural feedback (88.8\% of criteria under strict aggregation), but its reliability varies by dimension. CAPRA results are most dependable on well-defined extractive tasks, where presence/absence rests on objective evidence, and least so on interpretive criteria (A3, C1), where the lower inter-rater agreement reflects genuine subjectivity among raters rather than a pipeline failure (D1's $\kappa = 0$ being a known class-imbalance artifact; see Section~\ref{sec:rq1_results}). The highest pass rates fall on evidence-bound criteria such as actionable recommendations (C2), confirming that deterministic grounding is an effective way to curb LLM hallucinations in educational settings.}

\textbf{Implications for Educators.}
Our results suggest that CAPRA is best deployed as a \textit{first-pass teaching assistant} rather than a standalone grader. The 7.2--10.8$\times$ speedup enables instructors to offer rapid, formative feedback cycles, e.g., providing preliminary comments within hours of submission rather than weeks. {This also makes CAPRA suitable for intermediate evaluations of work in progress before the final submission. In this role, CAPRA can surface likely issues early, while human review concentrates on dimensions that require greater interpretive judgment, particularly \textit{Architecture \& Design Patterns} (A3) and \textit{Grounded Issues} (C1), where the evaluation task showed lower agreement among human raters.}

\textbf{Limitations.}
Several practical limitations should be noted. First, all student reports 
in our evaluation were written either in Italian or in English, while CAPRA's prompts and 
agent instructions are authored in English. Although \texttt{gpt-5.1} 
supports multilingual processing, we observed a concrete instance of 
language contamination that directly contributed to the single failure 
in the \textit{Template \& Tone} dimension (D1): one rater identified 
that a generated feedback report mixed Italian and English across 
different sections. Despite all agents being explicitly prompted to 
respond in English, prolonged exposure to a large Italian-language 
input document appears to have caused the model to drift toward the 
source language in isolated sections. While this was an isolated case, 
it highlights a subtle but real risk in cross-lingual deployments. 

Second, calibrating the scope and granularity of generated issues 
proved non-trivial. We observed a tension between overly generic 
feedback and production-oriented issues outside the pedagogical 
scope of an undergraduate course (e.g., security concerns or 
database transaction strategies). Resolving this currently requires 
iterative prompt refinement, undermining the plug-and-play 
usability of the system. A key open challenge is therefore 
enabling instructors to control feedback scope through 
high-level configuration rather than direct prompt engineering.

Third, the system currently depends on OpenAI's proprietary 
\texttt{gpt-5.1} model, introducing both a cost dependency 
($\sim$\$0.44/report) and a reproducibility constraint, as model 
behavior may change across API versions. Finally, the deterministic 
seed and temperature-zero settings mitigate but do not fully eliminate 
non-determinism in LLM outputs, as acknowledged by OpenAI's 
documentation.

\textbf{Comparison with Existing Approaches.}
Unlike ArTEMiS~\cite{krusche2018artemis}, which provides automated feedback primarily for code-based assignments through static analysis, CAPRA targets the complementary and underserved domain of open-ended architectural documentation. Compared to single-LLM approaches for essay and diagram grading~\cite{bouali2025grading,xie2024grade}, CAPRA's multi-agent decomposition with evidence anchoring provides two structural advantages: ({i})~{specialized agents help decompose the task into more focused checks}; and (ii)~the deterministic anchoring layer provides a verifiable trust boundary absent in end-to-end LLM evaluations.

\section{Threats to Validity}
We discuss the threats to validity of our study following established 
categorizations~\cite{runeson2009guidelines}.

\textbf{Construct Validity.}
The evaluation rubric (categories A1--D1) was designed by the authors to capture 
the key quality dimensions of architectural deliverables. This rubric may not cover 
all relevant aspects of software architecture assessment, potentially causing relevant 
quality dimensions to go unmeasured and leading to an incomplete picture of system 
performance. Furthermore, the binary Pass/Fail evaluation scale sacrifices 
{granularity compared to ordinal or continuous scales, which may obscure partial
correctness by forcing borderline cases into fully positive or fully negative
judgments.} As a mitigation, the rubric was grounded in
established software architecture quality criteria and refined through a pilot 
evaluation. Future work could adopt finer-grained rubrics (e.g., Likert scales) 
to capture partial compliance more accurately.

\textbf{Internal Validity.}
Both raters were authors of this paper, which introduces a potential evaluator bias 
toward favorable assessments of the system. This could have artificially inflated 
pass rates and inter-rater agreement. To mitigate this risk, the two raters 
performed their evaluations independently, without prior discussion, and Cohen's 
Kappa was adopted as the primary reliability metric to account for chance agreement. 
An external replication with independent raters would further strengthen confidence 
in the results.

Additionally, the Evidence Anchoring thresholds ($\tau_{min} = 0.45$, confidence 
filter at 0.65, trigram overlap at 0.27) were determined empirically through 
preliminary experiments. Different threshold configurations could yield different 
filtering behaviors, potentially retaining more hallucinated findings or discarding 
valid ones, thus directly affecting the pass rates reported for criteria such as C1. 
A systematic sensitivity analysis of these parameters is left as future work, as 
the current dataset size does not provide sufficient statistical power to derive 
robust optimal values.

Furthermore, the exclusive use of top-scoring reports introduces an additional 
threat to internal validity. This selection bias may have inflated pass rates and 
limited the discriminative power of the evaluation, as the system is never 
challenged with severely deficient deliverables, where hallucination rates and 
false positives could be substantially higher. This is a deliberate trade-off: low-scoring reports tend to contain pervasive
structural deficiencies that would hinder a focused assessment of subtle
architectural shortcomings. Extending the evaluation to the full spectrum of student
performance levels is planned as future work.

\textbf{External Validity.}
Our evaluation was conducted on 10 reports from a single software engineering course 
at one university, limiting generalization to other institutional contexts, course 
structures, and grading cultures. This could mean that the system's performance, 
and in particular the relevance of the automatically generated Knowledge Base 
features, may not transfer to courses with different architectural conventions or 
documentation standards. Furthermore, the system was evaluated using a single LLM configuration; performance may vary with different models, particularly 
open-source alternatives with potentially lower reasoning capabilities, which could 
reduce feedback quality and Evidence Anchoring accuracy. Replication across multiple 
institutions, languages, student performance levels, and model configurations is 
needed to establish broader validity, and is planned as a primary direction for 
future work.

\section{Replication Package}
\label{sec:replication}
To support reproducibility, a replication package is publicly available on Zenodo at \href{https://doi.org/10.5281/zenodo.20629900}{DOI: doi.org/10.5281/zenodo.20629900}. It contains: (i)~a self-contained, Java-based re-implementation of the CAPRA pipeline on the Spring AI framework~\cite{springai2024}, with the Python microservices for document parsing and multi-modal extraction; (ii)~the CSV export of the SWE features mined via the SALLMA workflow and HDBSCAN clustering for Knowledge Base generation; (iii)~the 10 CAPRA-generated PDF feedback reports, one per evaluated team (Teams 06, 12--20); (iv)~the full annotation spreadsheet with both raters' binary scores across the 8 criteria and 10 teams plus the evaluation taxonomy (Table~\ref{tab:taxonomy}), used for all inter-rater statistics in Section~\ref{sec:rq1_results}; and (v)~one original student deliverable used as input (shared with explicit consent), the remaining nine being withheld to protect student privacy, as public-release consent was not obtained.

\section{Conclusion}

This paper presented CAPRA, a multi-agent LLM system that automates formative feedback on software architecture deliverables. By orchestrating {multiple specialized agents} through a four-stage pipeline, CAPRA addresses the scalability bottleneck of manual review in software engineering education.

Our preliminary empirical viability assessment on 10 student reports demonstrated that CAPRA achieves moderate inter-rater agreement with human evaluators ($\kappa = 0.582$) and {satisfies 88.8\% of the evaluated criteria under strict aggregation}, while processing each report in {slightly over 4 minutes} compared to the 30--45 minutes required for manual review. The Evidence Anchoring mechanism proved particularly effective in grounding LLM findings and mitigating hallucinations. Meanwhile, the automated Knowledge Base generation pipeline enabled course-specific customization without manual rubric authoring.

\textbf{Future Work.}
Several directions emerge from this study. First, we plan to extend the evaluation to a larger and more diverse dataset spanning multiple universities, languages, and student performance levels. Second, we intend to investigate open-source LLM alternatives to reduce API cost dependencies and improve reproducibility. Third, integration with established learning management systems such as Moodle and ArTEMiS~\cite{krusche2018artemis} would enable seamless adoption in real classroom workflows. Finally, {since this work assessed the system strictly from a teacher's perspective, our next step is to directly evaluate its educational usefulness with students.}

\bibliographystyle{splncs04}
\bibliography{references}

@inproceedings{sallma2025,
  author    = {Marco Becattini and Roberto Verdecchia and Enrico Vicario},
  title     = {{SALLMA}: A Software Architecture for {LLM}-Based Multi-Agent Systems},
  booktitle = {Proceedings of the IEEE/ACM International Workshop on New Trends in Software Architecture (SATrends)},
  pages     = {5--8},
  year      = {2025},
  doi       = {10.1109/SATrends66715.2025.00006},
}

@article{petersen2023capstone, author = {Tenhunen, Saara and M\"{a}nnist\"{o}, Tomi and Luukkainen, Matti and Ihantola, Petri}, title = {A systematic literature review of capstone courses in software engineering}, year = {2023}, issue_date = {Jul 2023}, publisher = {Butterworth-Heinemann}, address = {USA}, volume = {159}, number = {C}, pages = {107191}, issn = {0950-5849}, doi = {10.1016/j.infsof.2023.107191}, journal = {Inf. Softw. Technol.}, month = jul, numpages = {21} }

@article{messer2024systematic,
  author  = {Marcus Messer and Neil C. C. Brown and Michael Kolling and Miaojing Shi},
  title   = {Automated grading and feedback tools for programming education: A systematic review},
  journal = {ACM Trans. Comput. Educ.},
  volume  = {24},
  number  = {1},
  pages   = {1--43},
  year    = {2024},
  doi     = {10.1145/3636515},
}

@Article{xu2025systematic,
AUTHOR = {Emirtekin, Emrah},
TITLE = {Large Language Model-Powered Automated Assessment: A Systematic Review},
JOURNAL = {Applied Sciences},
VOLUME = {15},
YEAR = {2025},
NUMBER = {10},
pages = {5683},
URL = {https://www.mdpi.com/2076-3417/15/10/5683},
ISSN = {2076-3417},
DOI = {10.3390/app15105683}
}

@inproceedings{xie2024grade,
author = {Xie, Wenjing and Niu, Juxin and Xue, Chun Jason and Guan, Nan},
title = {Grade Like a Human: Rethinking Automated Assessment with Large Language Models},
year = {2025},
isbn = {9798400722318},
publisher = {Association for Computing Machinery},
address = {New York, NY, USA},
doi = {10.1145/3769002.3769962},
booktitle = {Proceedings of the International Conference on Research in Adaptive and Convergent Systems},
pages = {1--8},
articleno = {21},
numpages = {8},
location = {
},
series = {RACS '25}
}

@inproceedings{guo2024multiagent,
  title     = {Large Language Model Based Multi-agents: A Survey of Progress and Challenges},
  author    = {Guo, Taicheng and Chen, Xiuying and Wang, Yaqi and Chang, Ruidi and Pei, Shichao and Chawla, Nitesh V. and Wiest, Olaf and Zhang, Xiangliang},
  booktitle = {Proceedings of the Thirty-Third International Joint Conference on
               Artificial Intelligence, {IJCAI-24}},
  publisher = {International Joint Conferences on Artificial Intelligence Organization},
  editor    = {Kate Larson},
  pages     = {8048--8057},
  year      = {2024},
  month     = {8},
  note      = {Survey Track},
  doi       = {10.24963/ijcai.2024/890},
}

@inproceedings{krusche2018artemis,
author = {Krusche, Stephan and Seitz, Andreas},
year = {2018},
month = feb,
pages = {284--289},
title = {ArTEMiS: An Automatic Assessment Management System for Interactive Learning},
booktitle = {Proceedings of the 49th ACM Technical Symposium on Computer Science Education},
doi = {10.1145/3159450.3159602}
}

@inproceedings{bouali2025grading,
title = {Toward Automated {UML} Diagram Assessment: Comparing {LLM}-Generated Scores with Teaching Assistants},
author = {Nacir Bouali and Marcus Gerhold and Tosif Ul Rehman and Faizan Ahmed},
year = {2025},
doi = {10.5220/0013481900003932},
volume = "1",
pages = {158--169},
booktitle = {Proceedings of the 17th International Conference on Computer Supported Education, CSEDU 2025},
}

@inproceedings{hong2024metagpt,
  author    = {Sirui Hong and Mingchen Zhuge and Jonathan Chen and Xiawu Zheng and Yuheng Cheng and Ceyao Zhang and Jinlin Wang and Zili Wang and Steven Ka Shing Yau and Zijuan Lin and Liyang Zhou and Chenyu Ran and Lingfeng Xiao and Chenglin Wu and J{\"u}rgen Schmidhuber},
  title     = {{MetaGPT}: Meta programming for a multi-agent collaborative framework},
  booktitle = {The Twelfth International Conference on Learning Representations},
  year      = {2024},
}

@article{wu2023autogen,
  author  = {Qingyun Wu and Gagan Bansal and Jieyu Zhang and Yiran Wu and Beibin Li and Erkang Zhu and Li Jiang and Xiaoyun Zhang and Shaokun Zhang and Jiale Liu and Ahmed Hassan Awadallah and Ryen W. White and Doug Burger and Chi Wang},
  title   = {{AutoGen}: Enabling next-gen {LLM} applications via multi-agent conversation},
  journal = {arXiv preprint arXiv:2308.08155},
  year    = {2023},
}

@inproceedings{qian2024chatdev,
  author    = {Chen Qian and Wei Liu and Hongzhang Liu and Nuo Chen and Yufan Dang and Jiahao Li and Cheng Yang and Weize Chen and Yusheng Su and Xin Cong and Juyuan Xu and Dahai Li and Zhiyuan Liu and Maosong Sun},
  title     = {{ChatDev}: Communicative agents for software development},
  booktitle = {Proceedings of the 62nd Annual Meeting of the Association for Computational Linguistics (Volume 1: Long Papers)},
  year      = {2024},
}

@article{li2024multiagent,

  author    = {Xinyi Li and Sai Wang and Siqi Zeng and Yu Wu and Yi Yang},
  title     = {A survey on {LLM}-based multi-agent systems: Workflow, infrastructure, and challenges},
  journal   = {Vicinagearth},
  volume    = {1},
  number    = {1},
  publisher = {Springer},
  year      = {2024},
  doi       = {10.1007/s44336-024-00009-2},
}

@inproceedings{chu2025agents,
    title = "{LLM} Agents for Education: Advances and Applications",
    author = "Chu, Zhendong  and
      Wang, Shen  and
      Xie, Jian  and
      Zhu, Tinghui  and
      Yan, Yibo  and
      Ye, Jingheng  and
      Zhong, Aoxiao  and
      Hu, Xuming  and
      Liang, Jing  and
      Yu, Philip S.  and
      Wen, Qingsong",
    editor = "Christodoulopoulos, Christos  and
      Chakraborty, Tanmoy  and
      Rose, Carolyn  and
      Peng, Violet",
    booktitle = "Findings of the Association for Computational Linguistics: EMNLP 2025",
    month = nov,
    year = "2025",
    address = "Suzhou, China",
    publisher = "Association for Computational Linguistics",
    doi = "10.18653/v1/2025.findings-emnlp.743",
    pages = "13782--13810",
    ISBN = "979-8-89176-335-7"
}

@inproceedings{zheng2023judging,
  author    = {Lianmin Zheng and Wei-Lin Chiang and Ying Sheng and Siyuan Zhuang and Zhanghao Wu and Yonghao Zhuang and Zi Lin and Zhuohan Li and Dacheng Li and Eric P. Xing and Hao Zhang and Joseph E. Gonzalez and Ion Stoica},
  title     = {Judging {LLM}-as-a-judge with {MT-Bench} and {Chatbot Arena}},
  booktitle = {Advances in Neural Information Processing Systems 36},
  volume    = {36},
  year      = {2023},
}

@article{dotsocr2025,
  author  = {Yumeng Li and Guang Yang and Hao Liu and Bowen Wang and Colin Zhang},
  title   = {{dots.ocr}: Multilingual Document Layout Parsing in a Single Vision-Language Model},
  journal = {arXiv preprint arXiv:2512.02498},
  year    = {2025},
}

@article{hattie2007feedback,
  author  = {John Hattie and Helen Timperley},
  title   = {The power of feedback},
  journal = {Review of Educational Research},
  volume  = {77},
  number  = {1},
  pages   = {81--112},
  year    = {2007},
}

@inproceedings{radermacher2013gaps,
  author    = {Alex Radermacher and Gursimran Walia},
  title     = {Gaps between industry expectations and the abilities of graduates},
  booktitle = {Proceedings of the 44th ACM Technical Symposium on Computer Science Education},
  pages     = {525--530},
  year      = {2013},
}

@article{keuning2019systematic,
  author  = {Hieke Keuning and Johan Jeuring and Bastiaan Heeren},
  title   = {A systematic literature review of automated feedback generation for programming exercises},
  journal = {ACM Trans. Comput. Educ.},
  volume  = {19},
  number  = {1},
  pages   = {1--43},
  year    = {2019},
}

@article{alamutka2005survey,
  author  = {Kirsti M. Ala-Mutka},
  title   = {A survey of automated assessment approaches for programming assignments},
  journal = {Computer Science Education},
  volume  = {15},
  number  = {2},
  pages   = {83--102},
  year    = {2005},
}

@inproceedings{ihantola2010review,
  author    = {Petri Ihantola and Tuukka Ahoniemi and Ville Karavirta and Otto Sepp{\"a}l{\"a}},
  title     = {Review of recent systems for automatic assessment of programming assignments},
  booktitle = {Proceedings of the 10th Koli Calling International Conference on Computing Education Research},
  pages     = {86--93},
  year      = {2010},
}

@inproceedings{singh2013automated,
  author    = {Rishabh Singh and Sumit Gulwani and Armando Solar-Lezama},
  title     = {Automated feedback generation for introductory programming assignments},
  booktitle = {Proceedings of the 34th ACM SIGPLAN Conference on Programming Language Design and Implementation},
  pages     = {15--26},
  year      = {2013},
}

@article{kasneci2023chatgpt,
  author  = {Enkelejda Kasneci and Kathrin Se{\ss}ler and Stefan K{\"u}chemann and Maria Bannert and Daryna Dementieva and Frank Fischer and Urs Gasser and Georg Groh and Stephan G{\"u}nnemann and Eyke H{\"u}llermeier and Stephan Krusche and Gitta Kutyniok and Tilman Michaeli and Claudia Nerdel and J{\"u}rgen Pfeffer and Oleksandra Poquet and Michael Sailer and Albrecht Schmidt and Tina Seidel and Matthias Stadler and Jochen Weller and Jochen Kuhn and Gjergji Kasneci},
  title   = {{ChatGPT} for good? {On} opportunities and challenges of large language models for education},
  journal = {Learning and Individual Differences},
  volume  = {103},
  pages   = {102274},
  year    = {2023},
}

@article{ji2023hallucination,
author = {Ji, Ziwei and Lee, Nayeon and Frieske, Rita and Yu, Tiezheng and Su, Dan and Xu, Yan and Ishii, Etsuko and Bang, Ye Jin and Madotto, Andrea and Fung, Pascale},
title = {Survey of Hallucination in Natural Language Generation},
year = {2023},
issue_date = {December 2023},
publisher = {Association for Computing Machinery},
address = {New York, NY, USA},
volume = {55},
number = {12},
pages = {1--38},
issn = {0360-0300},
doi = {10.1145/3571730},
journal = {ACM Comput. Surv.},
month = mar,
articleno = {248},
numpages = {38}
}

@article{huang2023hallucination,
author = {Huang, Lei and Yu, Weijiang and Ma, Weitao and Zhong, Weihong and Feng, Zhangyin and Wang, Haotian and Chen, Qianglong and Peng, Weihua and Feng, Xiaocheng and Qin, Bing and Liu, Ting},
title = {A Survey on Hallucination in Large Language Models: Principles, Taxonomy, Challenges, and Open Questions},
year = {2025},
issue_date = {March 2025},
publisher = {Association for Computing Machinery},
address = {New York, NY, USA},
volume = {43},
number = {2},
pages = {1--55},
issn = {1046-8188},
doi = {10.1145/3703155},
journal = {ACM Trans. Inf. Syst.},
month = jan,
articleno = {42},
numpages = {55}
}

@inproceedings{phung2023gpt4,
  author    = {Tung Phung and Jos{\'e} Cambronero and Sumit Gulwani and Tobias Kohn and Rupak Majumdar and Adish Singla and Gustavo Soares},
  title     = {Generating high-precision feedback for programming syntax errors using large language models},
  booktitle = {Proceedings of the 16th International Conference on Educational Data Mining},
  year      = {2023},
}

@inproceedings{li2023camel,
  author    = {Guohao Li and Hasan Abed Al Kader Hammoud and Hani Itani and Dmitrii Khizbullin and Bernard Ghanem},
  title     = {{CAMEL}: Communicative agents for ``mind'' exploration of large language model society},
  booktitle = {Advances in Neural Information Processing Systems 36},
  volume    = {36},
  year      = {2023},
}

@inproceedings{chen2024agentverse,
  author    = {Weize Chen and Yusheng Su and Jingwei Zuo and Cheng Yang and Chenfei Yuan and Chi-Min Chan and Heyang Yu and Yaxi Lu and Yi-Hsin Hung and Chen Qian and Yifan Qin and Xin Cong and Ruobing Xie and Zhiyuan Liu and Maosong Sun and Jie Zhou},
  title     = {{AgentVerse}: Facilitating multi-agent collaboration and exploring emergent behaviors},
  booktitle = {The Twelfth International Conference on Learning Representations},
  year      = {2024},
}

@inproceedings{wang2024faireval,
  author    = {Peiyi Wang and Lei Li and Liang Chen and Zefan Cai and Dawei Zhu and Binghuai Lin and Yunbo Cao and Lingpeng Kong and Qi Liu and Tianyu Liu and Zhifang Sui},
  title     = {Large language models are not fair evaluators},
  booktitle = {Proceedings of the 62nd Annual Meeting of the Association for Computational Linguistics (Volume 1: Long Papers)},
  year      = {2024},
}

@inproceedings{liu2023geval,
  author    = {Yang Liu and Dan Iter and Yichong Xu and Shuohang Wang and Ruochen Xu and Chenguang Zhu},
  title     = {{G-Eval}: {NLG} evaluation using {GPT-4} with better human alignment},
  booktitle = {Proceedings of the 2023 Conference on Empirical Methods in Natural Language Processing},
  year      = {2023},
}

@inproceedings{kim2024prometheus,
  author    = {Seungone Kim and Jamin Shin and Yejin Cho and Joel Jang and Shayne Longpre and Hwaran Lee and Sangdoo Yun and Seongjin Shin and Sungdong Kim and James Thorne and Minjoon Seo},
  title     = {Prometheus: Inducing fine-grained evaluation capability in language models},
  booktitle = {The Twelfth International Conference on Learning Representations},
  year      = {2024},
}

@misc{openai2024gpt4o,
  author       = {{OpenAI}},
  title        = {Hello {GPT-4o}},
  year         = {2024},
  howpublished = {\url{https://openai.com/index/hello-gpt-4o/}},
  note         = {Accessed: June 2026}
}

@article{cohen1960kappa,
  author  = {Jacob Cohen},
  title   = {A coefficient of agreement for nominal scales},
  journal = {Educational and Psychological Measurement},
  volume  = {20},
  number  = {1},
  pages   = {37--46},
  year    = {1960},
}

@article{landis1977kappa,
  author  = {J. Richard Landis and Gary G. Koch},
  title   = {The measurement of observer agreement for categorical data},
  journal = {Biometrics},
  volume  = {33},
  number  = {1},
  pages   = {159--174},
  year    = {1977},
}

@article{kitchenham2002guidelines,
  author  = {Barbara A. Kitchenham and Shari Lawrence Pfleeger and Lesley Pickard and Peter W. Jones and David C. Hoaglin and Khaled El~Emam and Jarrett Rosenberg},
  title   = {Preliminary guidelines for empirical research in software engineering},
  journal = {IEEE Trans. Softw. Eng.},
  volume  = {28},
  number  = {8},
  pages   = {721--734},
  year    = {2002},
}

@article{runeson2009guidelines,
  author  = {Per Runeson and Martin H{\"o}st},
  title   = {Guidelines for conducting and reporting case study research in software engineering},
  journal = {Empirical Software Engineering},
  volume  = {14},
  number  = {2},
  pages   = {131--164},
  year    = {2009},
}

@inproceedings{wang2023selfconsistency,
  author    = {Xuezhi Wang and Jason Wei and Dale Schuurmans and Quoc V. Le and Ed H. Chi and Sharan Narang and Aakanksha Chowdhery and Denny Zhou},
  title     = {Self-consistency improves chain of thought reasoning in language models},
  booktitle = {The Eleventh International Conference on Learning Representations},
  year      = {2023},
}

@article{levenshtein1966binary,
  author  = {Vladimir I. Levenshtein},
  title   = {Binary codes capable of correcting deletions, insertions, and reversals},
  journal = {Soviet Physics Doklady},
  volume  = {10},
  pages   = {707--710},
  year    = {1966},
}

@inproceedings{brown2020language,
  author    = {Tom B. Brown and Benjamin Mann and Nick Ryder and Melanie Subbiah and Jared D. Kaplan and Prafulla Dhariwal and others},
  title     = {Language models are few-shot learners},
  booktitle = {Advances in Neural Information Processing Systems 33},
  volume    = {33},
  pages     = {1877--1901},
  year      = {2020},
}

@article{fan2023llmse,
  author  = {Hou, Xinyi and Zhao, Yanjie and Liu, Yue and Yang, Zhou and Wang, Kailong and Li, Li and Luo, Xiapu and Lo, David and Grundy, John and Wang, Haoyu},
  title   = {Large Language Models for Software Engineering: A Systematic Literature Review},
  journal = {ACM Transactions on Software Engineering and Methodology},
  volume  = {33},
  number  = {8},
  pages   = {1--79},
  year    = {2024},
  doi     = {10.1145/3695988},
  note    = {Article 220}
}

@misc{pymupdf2024,
  author       = {{Artifex Software, Inc.}},
  title        = {{PyMuPDF}: Python Bindings for {MuPDF}},
  year         = {2024},
  url          = {https://pymupdf.readthedocs.io},
  note         = {Version 1.24, accessed 2024}
}

@misc{springai2024,
  author       = {{VMware, Inc. and Spring Contributors}},
  title        = {Spring {AI}: An Application Framework for {AI} Engineering},
  year         = {2024},
  howpublished = {\url{https://spring.io/projects/spring-ai}},
  note         = {Version 1.0.x. Accessed: February 2026}
}

@inproceedings{campello2013hdbscan,
  author    = {Campello, Ricardo J. G. B. and Moulavi, Davoud and Sander, J{\"o}rg},
  title     = {Density-Based Clustering Based on Hierarchical Density Estimates},
  booktitle = {Advances in Knowledge Discovery and Data Mining -- PAKDD 2013},
  year      = {2013},
  pages     = {160--172},
  publisher = {Springer},
  series    = {Lecture Notes in Computer Science},
  volume    = {7819},
  doi       = {10.1007/978-3-642-37456-2_14}
}

@article{normalizedlevenshtein2007,
  author={Yujian, Li and Bo, Liu},
  journal={IEEE Transactions on Pattern Analysis and Machine Intelligence}, 
  title={A Normalized Levenshtein Distance Metric}, 
  year={2007},
  volume={29},
  number={6},
  pages={1091-1095},
  doi={10.1109/TPAMI.2007.1078}}

@article{approxmatching2001,
author = {Navarro, Gonzalo},
title = {A guided tour to approximate string matching},
year = {2001},
issue_date = {March 2001},
publisher = {Association for Computing Machinery},
address = {New York, NY, USA},
volume = {33},
number = {1},
issn = {0360-0300},
doi = {10.1145/375360.375365},
journal = {ACM Comput. Surv.},
month = mar,
pages = {31--88},
numpages = {58}
}

@ARTICLE{marzal1993computation,
  author={Marzal, A. and Vidal, E.},
  journal={IEEE Transactions on Pattern Analysis and Machine Intelligence}, 
  title={Computation of normalized edit distance and applications}, 
  year={1993},
  volume={15},
  number={9},
  pages={926-932},
  doi={10.1109/34.232078}}

\end{document}